\documentclass[aps,preprint,amsmath,floatfix,prb]{revtex4}
\usepackage{graphicx}
\usepackage{graphics}
\usepackage{epsf}
\usepackage{bm}
\begin{document}
\flushbottom

\title{Structural, electronic, and magnetic properties of Mn-doped Ge nanowires by $ab\;initio$ calculations}

\author{J. T. Arantes, Ant\^{o}nio J. R. da Silva, and A. Fazzio}

\affiliation{ Instituto de F\'\i sica, Universidade de S\~ao
Paulo, CP 66318, 05315-970 S\~ao Paulo, SP, Brazil}

\date{\today}

\begin{abstract}
Using {\it ab initio} total  energy density-functional theory calculations, we investigate the electronic,
structural and magnetic properties of  Manganese doped Germanium  nanowires.
The nanowires have been constructed along the [110] direction and the dangling bonds on the
surface have been saturated by hydrogen atoms. We observed that the    
Mn has lower formation energy at the center of the wire when compared to regions close to the
surface.  
The Mn-Mn coupling has lower energy for a high-spin configuration except  when they are first 
nearest neighbors. 
These results show that Ge:Mn nanowires are potential candidates for ferromagnetic 
quasi-one dimensional systems.
\end{abstract}

%\pacs{}

\maketitle

\section{Introduction}

Semiconductor nanowires \cite{pyang05} are considered by many researchers as the class of nanomaterials that will become the major player in the future electronic technology. Among these nanowires, those composed by silicon (SiNWs) and germanium (GeNWs) have an intrinsic interest\cite{{wu04},lauhon02}  because they may be more easily integrated with the current silicon-based technology. Besides the search for nanomaterials, a field that has attracted a great deal of attention due to possible developments of new devices is the study of diluted magnetic semiconductors (DMS)\cite{igor,natmater}. The DMS are usually composed by transition metals embedded in a semiconductor matrix, in a high enough concentration to render the material ferromagnetic below a certain critical temperature Tc. The highest critical temperatures so far achieved have been on III-V semiconductors. There is, however, a great interest in obtaining a DMS based on a type-IV semiconductor. Reports of ferromagnetism in Mn$_x$Ge$_{1-x}$ have been recently made\cite{park01,scho}, as well as in Mn-implanted silicon \cite{bolduc}. There is an important difference between Mn doping in Si and Ge. A Mn impurity favors an interstitial site in Si, whereas in Ge it prefers a substitutional site \cite{prb04}. As a consequence, a Mn substitutional in Ge cannot diffuse as easily as an interstitial Mn in Si\cite{naka91}, allowing the introduction of a larger number of Mn in Ge without their diffusion and subsequent clustering.

Considering what has been described above, it seems relevant to study the behavior of Mn-doped Ge nanowires, and to investigate the effect of the quantum confinement on the properties of these DMS nanowires. The magnetic ordering in Mn-doped nanoparticles and nanowires semiconductors has been recently addressed by numerous investigations \cite{poddar,schmidt,chelik05}. In particular, LiberÕs group has obtained a general synthesis procedure of Mn-doped semiconductor nanowires\cite{rado05}. These controlled processes to obtain nanowires doped with magnetic ions give us opportunities to understand the magnetic order in low-dimensional materials for future applications in spintronic nano-devices.

In this work we study, from theoretical point of view, the structural, electronic and magnetic properties of Mn doped GeNWs. In particular, we observe that the Mn has a lower formation energy at the center of the wire when compared to regions close to the surface. In the next section we describe the procedure we used in our calculations, followed by the section with the results and discussion. Finally, the main conclusions are summarized in the last section.

\begin{figure}[ht]
\includegraphics[width= 8.5cm]{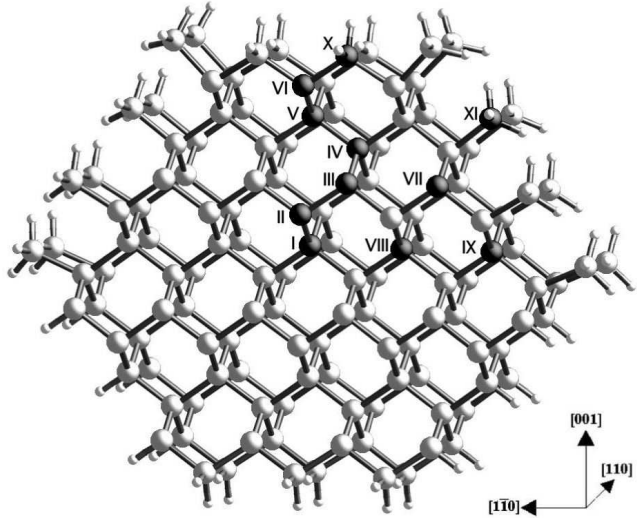}
\caption{\label{struct} Atomic geometry for the Ge nanowire with  diameter   $d~\simeq~27.0$~{\AA}. The larger gray spheres represent the Ge atoms and the small ones represent the hydrogen atoms at the surface. The black spheres  represent the studied substitutional positions of the Mn impurity.}
\end{figure}

\section{Calculation Procedure}

All our results are obtained using  total energy {\it ab initio}
calculations based on spin-polarized density functional theory
within the generalized gradient approximation (GGA)
\cite{perdew92} for the exchange-correlation potential. We used ultrasoft pseudopotentials
\cite{vanderbilt90} and a plane wave expansion up to 230.0 eV, as
   implemented in the VASP code \cite{kresse93}.
For bulk calculations we  used  64 Ge atoms in a  cubic supercell
with a ($3\times3\times3$) Monkhorst-Pack Brillouin zone sampling.
We  studied  four different NWs   with 
zinc-blend structure  grown  along  the [110] direction.
The NW with  diameter  $d \simeq 35.0$~ {\AA}  contains 322 atoms, with 218 Ge atoms and 
saturated by 104 H atoms. For the $d \simeq 27.0$~{\AA} NW, used  for most of our 
studies of Mn doping (figure~\ref{struct}),
we have  214 atoms with 134 Ge  and 80 H atoms. 
The  diameter $d \simeq 18.7$~{\AA}  contains  116 atoms  with 76 Ge and 40 H atoms.
Finally, for the small diameter ($d \simeq 14.7$~{\AA}) the total number of  atoms  used was 60 with 36 Ge and 24 H atoms.
For the nanowire
calculations, we used a Brillouin zone sampling of three  {\it
k}-points corresponding to a ($1\times1\times3$) Monkhorst-Pack
grid.  In all calculations the positions
of all atoms in the supercell were allowed to relax until all the
\begin{figure}[ht]
\includegraphics[width= 8.5cm]{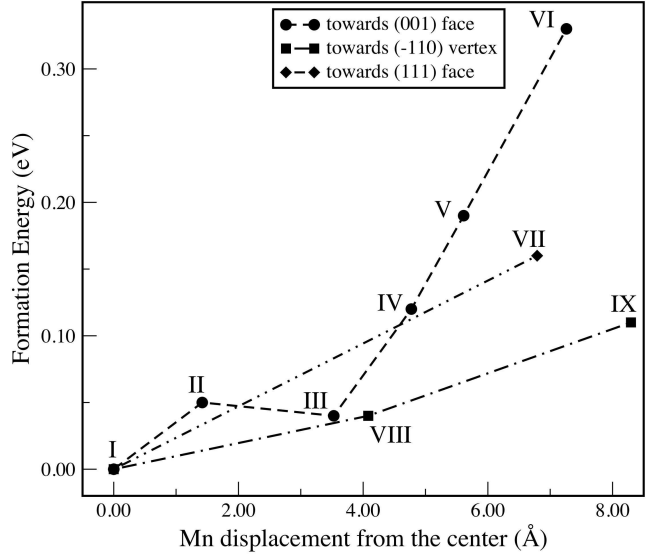}
\caption{\label{eform} Total energy for the  substitutional Mn dopant sites in the Ge
nanowires as a function of the dopant distance from the center of the nanowire. The reference
zero energy is for the Mn in the center of the wire. The dashed lines are only  guides to the eye, and they connect sites along  paths towards either a particular face or a vertex.}
\end{figure} 
forces were smaller than 0.02 eV/{\AA}.
For all impurity
calculations we have not fixed the value of the total spin of the
supercell. 
In each case, however, we used as an initial guess a Mn
high-spin configuration. 

\section{Results and Discussions}

All the GeNWs  calculated with different diameters, as described below,  have a hexagonal 
cross-section shape that exposes  \{111\} and  \{001\} faces (see figure \ref{struct}). The
corresponding angle between 
the exposed  \{001\} and  \{111\} (\{1$\overline{1}$1\}) face is 54.92$^o$ (54.91$^o$).
 The  angle between the \{111\} face and  the \{$\overline{1}1\overline{1}$\} 
face is 70.74$^o$, in agreement  with experimental results \cite{tobias05}.
The average Ge-Ge bond length is 2.49~{\AA} close to the center of the wire
and 2.47~{\AA} around the surface.
The Ge-H bond length in the surface  has an average of 1.54~{\AA}.
For the pure nanowire, as expected, the band gap increases with the quantum confinement. The calculated band gaps are 0.53, 0.78, 1.00
and 1.34 eV for the diameters $\simeq$35.0, $\simeq$27.0, $\simeq$18.7, and $\simeq$14.7~{\AA} respectively \cite{dft00}.

In figure \ref{eform}  we show our results for  the substitutional Mn formation energy for the non-equivalent sites  shown
in figure \ref{struct} .
The energy reference is taken for  the Mn substitutional in the center of the nanowire. The dashed lines are only  guides to the eye, and they connect sites
along  paths towards either a particular face or a vertex. The formation energy, $E_{f}^{S}$, is given by
\begin{equation}
E_{f}^{S} = (E_{def} + \mu_{Ge}) - E_{nw} - \mu_{Mn},
\label{eq2}
\end{equation}
\noindent where $E_{def}$ is the total energy of the supercell
with the Mn atom at a substitutional site and $E_{nw}$ is the total
energy of the nanowire without the Mn atom. The
$\mu_{Ge}$  and $\mu_{Mn}$ are  Ge and Mn  chemical potentials, respectively.
\begin{table}[ht]
\centering
\caption{\label{dist} The Mn-Ge bond length. The first column
represents the labeled (I to XI) configurations of figure \ref{struct}. In the second (third) column we show 
the longitudinal (cross-sectional) distances.
 }
\begin{tabular}{ccc} \hline \hline
\multicolumn{1}{c}{Config.}&
\multicolumn{1}{c}{Long. distances (\AA) }&
\multicolumn{1}{c}{Cross distances (\AA)}\\
\hline
I              &  2.46, 2.46 & 2.48, 2.48  \\
II              &  2.46, 2.46 & 2.48, 2.49  \\
III              & 2.47, 2.46 &  2.48, 2.48 \\
IV              &  2.45, 2.46 & 2.48, 2.48  \\
V              &  2.44, 2.46 & 2.47, 2.47  \\
VI              &  2.54, 2.49 & 2.46, 2.47  \\
VII              &  2.47, 2.47 & 2.49, 2.49  \\
VIII              & 2.46, 2.46  & 2.47, 2.48  \\
IX              &  2.49, 2.48 & 2.46, 2.49 \\
X              &   2.55 &  2.49, 2.52 \\
XI              &  2.56 & 2.55 \\
\hline \hline
\end{tabular}
\end{table}
The results show that Mn has lower 
 formation energy at the center of the Ge-nanowire. 

In table \ref{dist} we show the distances between the Mn and its nearest neighbor Ge atoms
for the sites labeled in figure \ref{struct}.
There are two types of neighbors and therefore two types of distances; the ones oriented along the
direction of growth of the nanowire, which  we will call longitudinal distances, and the ones oriented
perpendicular to the growth direction, which we call cross-sectional distances.
For Mn at the positions I to VIII, except position VI,  there is a small distortion, presenting a slightly distorted local
T$_d$ symmetry. In all these configurations the Mn atom is surrounded by four
Ge neighbors not bonded to saturating  H-atoms. In general, the Mn nearest neighbors   
suffer a bond length decrease of  $0.8\!\%$  up to $1.6\!\%$ when compared to a pure germanium nanowire.
The position VI and IX  present  a local C$_{3v}$ symmetry. These positions are at sub-surface layers,
allowing a larger relaxation for the Mn impurity.

\begin{figure*}[ht]
\includegraphics[angle=0,width=16.5cm]{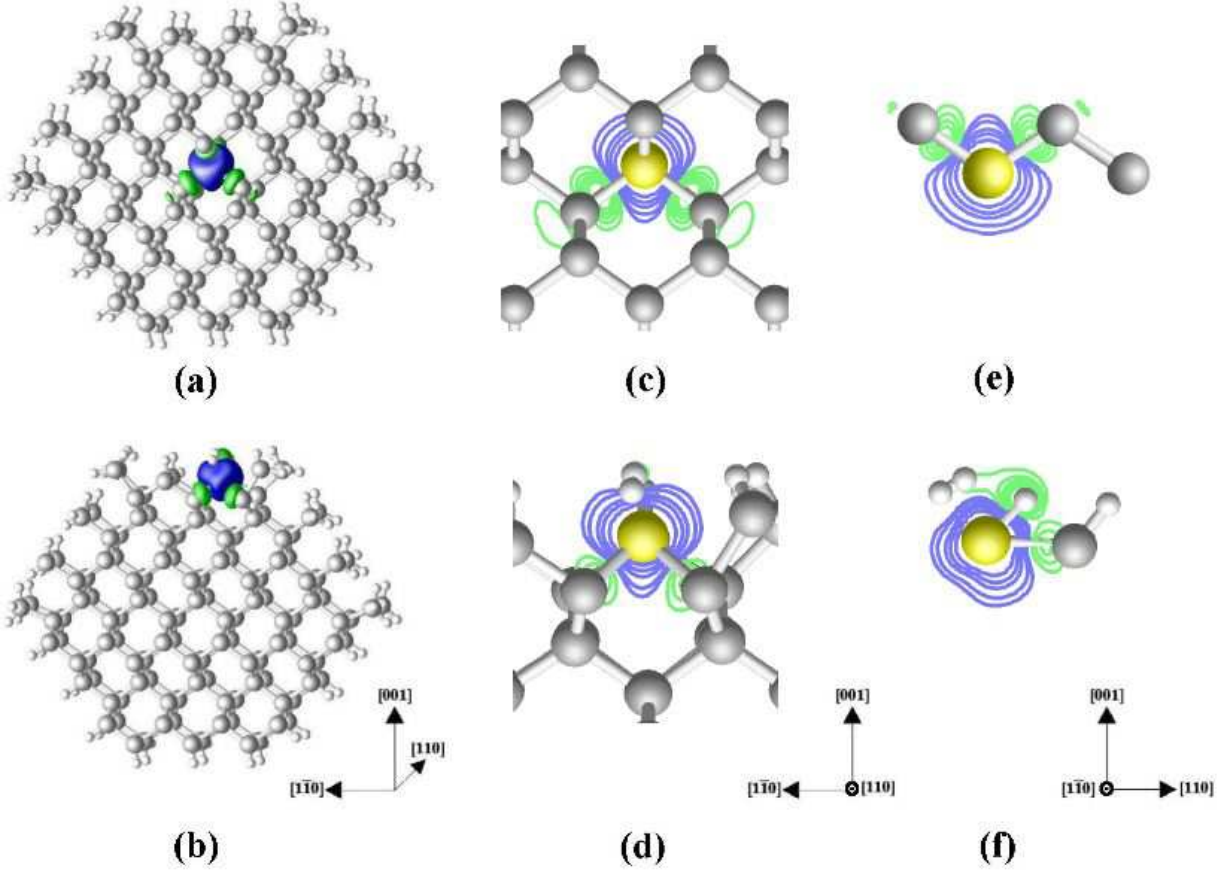}
\caption{(Color online) Isosurface for the net magnetization $m({\bf r})=\rho_{up}({\bf r}) - \rho_{down}({\bf r})$ for a) Mn atom at position I,  b) Mn atom at position X. Blue (green) regions represent predominant 
$\rho_{up}$ $(\rho_{down})$ electronic densities. The more spherical isosurfaces correspond to a net 
spin value of $+0.01e/${\AA}$^3$ and the four $p$-orbitals shaped isosurfaces correspond to a net 
spin value of $-0.01e/${\AA}$^3$. In (c) and (d) we show the cross-sectional views and in (e) an (f)
are the longitudinal views related to the isosurfaces depicted in (a) and (b). The outermost (innermost)
green line has $m({\bf r})=-0.01$ ($m({\bf r})=-0.05$). The innermost (outermost) blue line has
$m({\bf r})=0.20$ ($m({\bf r})=0.01$). }
\label{netmag}
\end{figure*}

We observed a huge structural change when the  substitutional Mn is at the surface of the nanowire, 
 at positions X and XI.  The Mn atom makes a bond with the Ge second
 neighbor from the  \{001\} and  \{111\}  faces for position X and XI respectively. The Mn atom becomes bonded to three Ge atoms and one H atom for position X and bonded to two Ge atoms and two H atoms for position XI. Consequently, there is a reaction with the formation of an almost  free-H$_2$ molecule. In this configuration, the impurity 
presented a lower
formation energy,  around 0.6 eV below  site-I. 
This indicates that once Mn atoms are placed inside the 
NWs, there will be an energetic barrier opposing its migration towards the surface.
However, for H-saturated wires, there might also be a large concentration of Mn atoms at the surface.
We also looked for Mn at a tetrahedral interstitial  site,
because  in the bulk the Mn at substitutional position  has lower formation energy compared with interstitial
one. The difference between the interstitial and substitutional formation energies in the center of the
nanowire are 0.69 eV and 0.78 eV for diameters 27.0 {\AA} and  18.7 {\AA} respectively. 

In figure \ref{netmag} we show
the net local magnetization for Mn, $m({\bf r})=\rho_{up}({\bf r}) - \rho_{down}({\bf r})$,
  a) at position I  and b) Mn at position X.
The behavior of the net local magnetization  is highly localized around the Mn atom
 with an opposite magnetization that comes from $p$-states contribution from Ge nearest-neighbors,
 which is similar to what is observed for Mn in bulk\cite{dalpian03B,ajrsilva00}. 
The integrated value of $m({\bf r})$ is about 3.00 $\mu_B$ for all cases studied.
The exception was for Mn at position VI where there is an increase of the magnetization to
3.66 $\mu_B$. 

\begin{figure}[!h]
\includegraphics[angle=0,width=5.5cm]{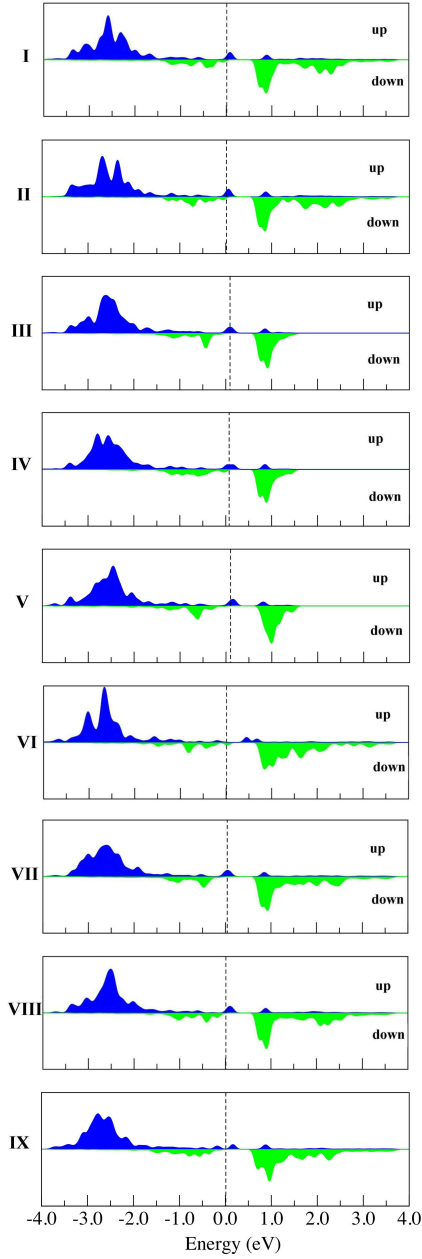}
\caption{ (Color online) Mn impurity partial spin  DOS, projected at the $\Gamma$-point,  for  Mn
at different substitutional sites, (see figure~\ref{struct}). The dashed lines represent  the Fermi energy.}
\label{dos}
\end{figure}

\begin{table}[ht]
\centering
\caption{\label{acopla} The magnetic  coupling between Mn-Mn atoms. The first column
represents the labeled (I to XI) configurations of figure \ref{struct}. The  d$_{Mn-Mn}$ shows the distance
between the two Mn atoms. The $\Delta E_f^{(AFM-FM)}$ is the total energy  difference between the 
AFM and FM states. The next two columns represent the local magnetic moment at each Mn atom. The
 last column gives the Mn-Mn direction.
 The first two lines represent interactions between Mn oriented mostly along the longitudinal direction,
 whereas the others are interactions between Mn atoms positioned mostly along the same cross-section.}
\begin{tabular}{cccccc} \hline \hline
\multicolumn{1}{c}{Configuration}&
\multicolumn{1}{c}{d$_{Mn-Mn}$ }&
\multicolumn{1}{c}{$\Delta E_f^{(AFM-FM)}$}&
\multicolumn{1}{c}{$\mu_1$(I)}&
\multicolumn{1}{c}{$\mu_2$}&
\multicolumn{1}{c}{direction}\\
\multicolumn{1}{c}{}&
\multicolumn{1}{c}{({\AA}) }&
\multicolumn{1}{c}{(eV)}&
\multicolumn{1}{c}{($\mu_B$)}&
\multicolumn{1}{c}{($\mu_B$)}\\
\hline
I + I                & 4.07 & 0.22 &  3.48 & 3.48 & [110]\\
I + I $^\text{{\cite{extra01}}}$& 8.14 & 0.14 & 3.45 & 3.45 & [110]\\ \hline
I + 1$^{st}$  & 2.00 & -0.69&-2.87 & 2.87 & [$\overline{1}$1$\overline{1}$]\\
I + VIII               & 4.03 & 0.22 &  3.47 & 3.47 & [$\overline{1}$10]\\
I + IV              & 4.79   & 0.12  & 3.50 & 3.46 & [3$\overline{2}$0]\\
I + V               & 5.70 & 0.07 & 3.49 & 3.44 & [001]\\ 
I + VII             & 7.13 & 0.07 &  3.52 & 3.44 & [$\overline{3}$62]\\
I + VI             & 7.61 & 0.04 &  3.60 & 3.38 & [$\overline{4}$01]\\
I + IX             & 8.33 & 0.10 &  3.42 & 3.54 & [$\overline{1}$10]\\ 
I + X             & 8.89 & 0.00 &  3.16 & 3.46 & [$\overline{6}$2$\overline{1}$]\\
\hline \hline
\end{tabular}
\end{table}

In figure \ref{dos} we plot the Mn-$d$ partial spin density of states (PDOS), projected at the
$\Gamma$-point,  for the substitutional Mn atom at  different sites in the GeNWs, (see figure \ref{struct}). For all  positions, similarly
 to Ge-bulk, the Mn atom introduces  majority spin levels with strong $d$ character and which are
 resonant with the 
 valence band (VB). The highest occupied orbital (HOMO) and   the lowest unoccupied  orbital (LUMO)
  in the gap region have a strong $p$-character from Ge-atoms and a small $d$-component from Mn.

\begin{figure}[ht]
\includegraphics[width= 8cm]{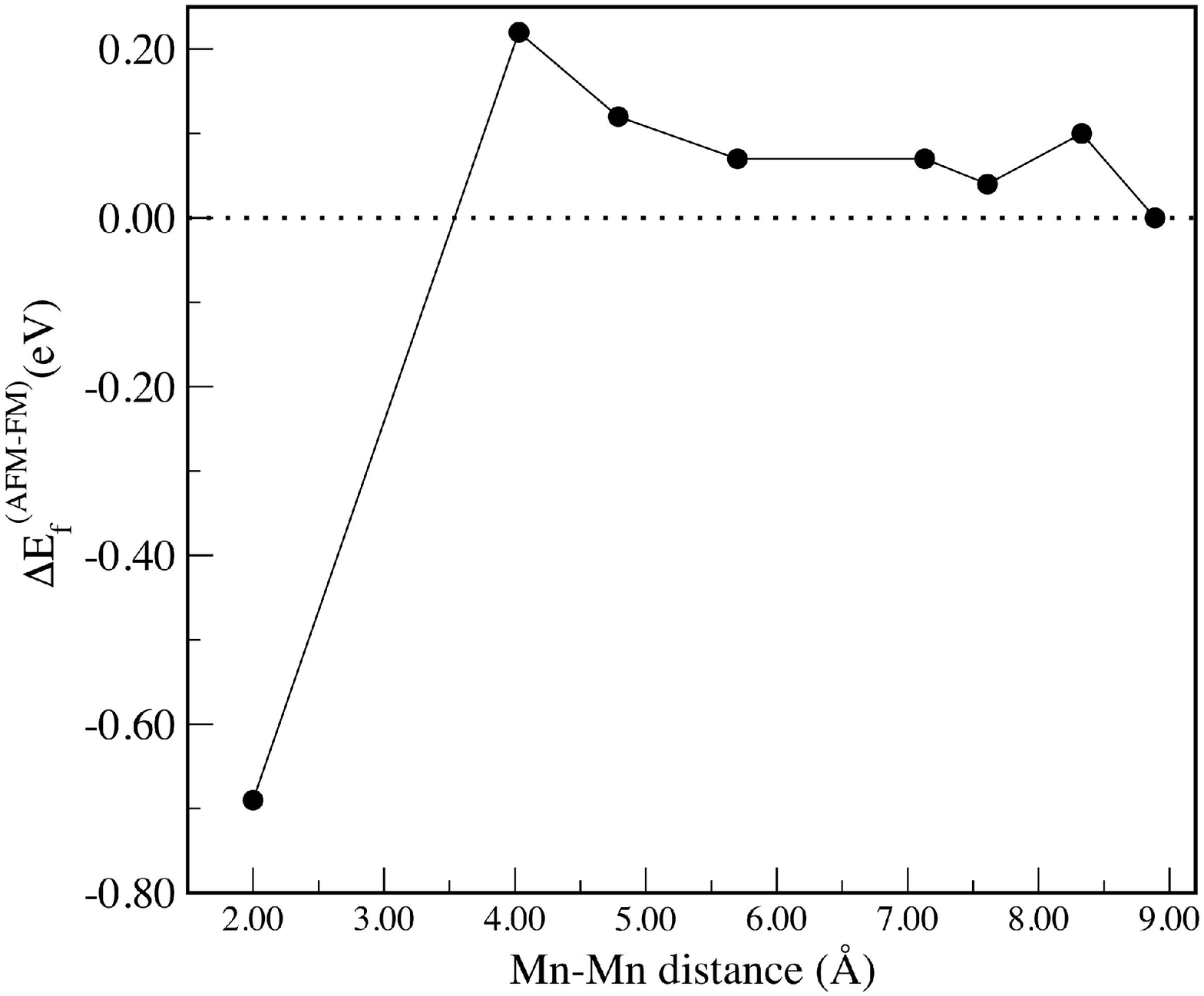}
\caption{\label{difener} Total energy difference between AFM and FM states, for  
Mn-Mn coupling as a function of the Mn-Mn distance. 
They correspond to the last eight lines in table \ref{acopla}. The dashed line is only  guides to the eye.}
\end{figure}

\begin{figure}[ht]
\includegraphics[angle=0,width=8.5cm]{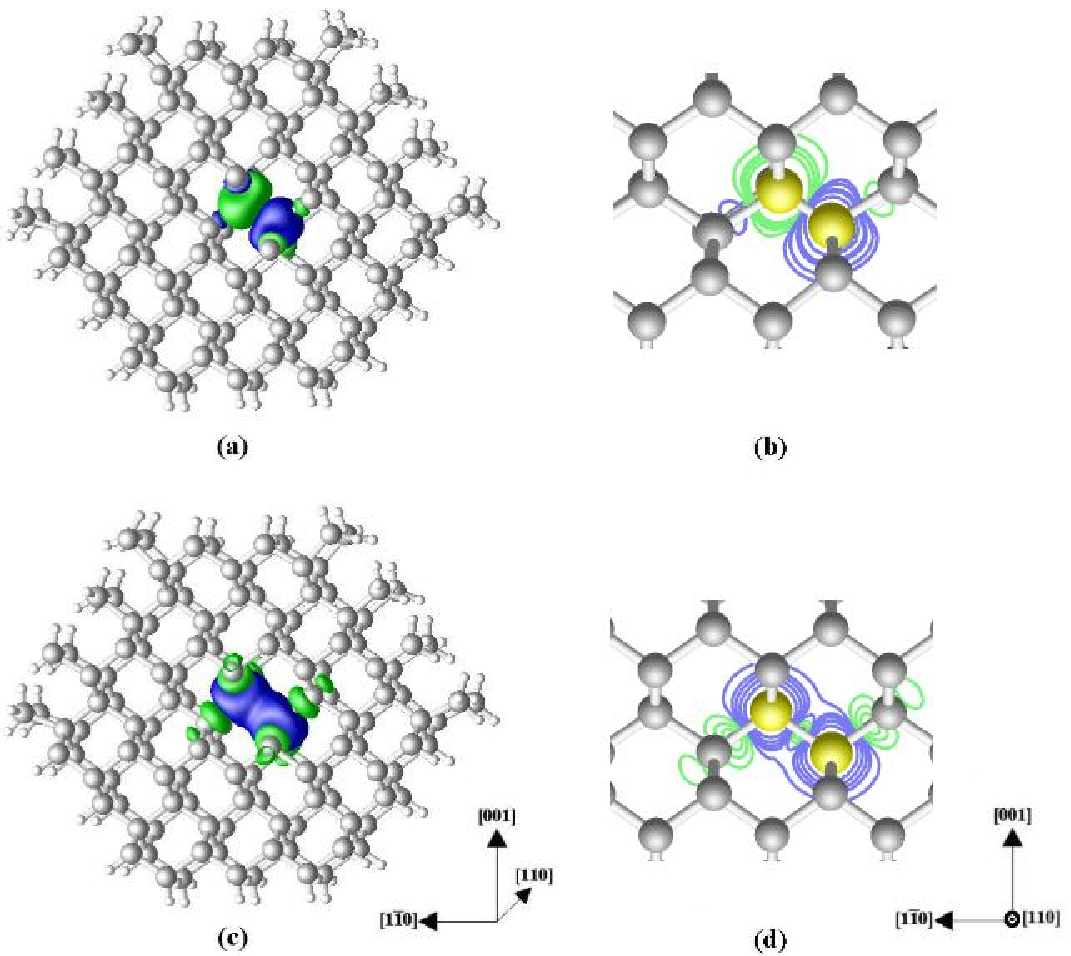}
\caption{(Color online) Isosurfaces for the local magnetization $m({\bf r})=\rho_{up}({\bf r}) - \rho_{down}({\bf r})$ for two Mn atoms, one at position I and another as  its  first neighbor. The small spheres represent the hydrogen atoms, the gray ones represent the Ge atoms and the yellow
(large) spheres  represent the Mn atoms.  Blue (green) regions show predominant $\rho_{up}$ $(\rho_{down})$ electronic densities. The more spherical isosurfaces correspond to a net spin value of 
$+0.01e/${\AA}$^3$ (blue) and $-0.01e/${\AA}$^3$ (green). In (a) and (b) we present an AFM state and in (c) and (d) a FM state. The most stable configuration, for this position,  is an antiferromagnetic state.
In (b) and (d) we show the contour plots (in $e/${\AA}$^3$)  for the local magnetization at a plane, that
passes through the Mn atoms.  The green lines correspond to the isosurfaces -0.01, -0.03, -0.05, -0.07
and -0.09, whereas the blue ones correspond to the values 0.01, 0.03, 0.05, 0.07 and 0.09.}
\label{netmag2}
\end{figure}

To understand the possibility of magnetic ordering, we studied the  coupling between  Mn-Mn atoms.  We  replaced two
Ge-atoms by Mn and  enforced different  total spin configurations .
For each of these Mn-pairs we computed the total energies for high spin state (ferromagnetic-FM)
and low spin state (antiferromagnetic-AFM).
Our simulation corresponds to a Mn concentration close to 2.0~$\%$.
We choose a few  different pair  configurations and the results are summarized in Table \ref{acopla}. 
The notation in the first column of the table follow the labeled (I to XI) configurations presented in
figure~\ref{struct}. 
The first two lines represent interactions between Mn oriented mostly along the longitudinal direction,
 whereas the others are interactions between Mn atoms positioned mostly along the same cross-section.
For example, the I~+~VI Mn-Mn pair means one Mn atom at site I and another at site VI. In the second
column we present the Mn-Mn distance. In the third column we show the total energy difference
($\Delta E_f$) between the antiferromagnetic (AFM) and ferromagnetic (FM) ground states.
The next two columns represent the local magnetic moments at each one of the Mn atoms where $\mu_1(I)$ is
the local magnetic moment for the Mn at  position I  and $\mu_2$ 
represents  the local magnetic moment for the second Mn atom. The last column gives the Mn-Mn
direction. 
For the Mn atom at position I and another Mn as its first neighbor, we found an 
antiferromagnetic ground state with an energy difference $\Delta$E$_f^{(AFM-FM)}$~=~-0.69~eV, as 
similarly  to what has been observed in bulk calculations\cite{stroppa03}.
Position I~+~VII  has a Mn-Mn distance smaller  than position  I~+~IX. However, the energy difference between AFM and FM states is slightly  larger in  I~+~IX.
This is caused by variations of the coupling with the crystallographic directions\cite{ajrsilva00,zhao05}.
For example, the I-IX Mn belong to the same Mn-Ge-Ge-Ge-Mn ``zig-zag chain" perpendicular to the
growth direction.  
For one Mn at position I, in the center,  and another at position X, at the surface, with a large distance
$\sim$~9~{\AA}, the high-spin  and low-spin configurations are degenerate.

Comparing the results for different Mn-Mn distances in  the GeNWs, see figure \ref{difener}, we do not see 
an oscillatory AFM-FM behavior  in  the wire as has been observed  for Ge-bulk \cite{stroppa03}.
We observe that  for Mn-Mn distances ranging  between 4.00~{\AA} to 8.00~{\AA}  the ferromagnetic
coupling is very similar.

In figure~\ref{netmag2} we show, for two Mn atoms as first neighbors,  
the local magnetization, $m({\bf r})=\rho_{up}({\bf r}) - \rho_{down}({\bf r})$,  and its contour plots 
(in $e/${\AA}$^3$).
The cross-sectional plane  passes through the Mn atoms. 
The Mn atoms are  with spin flipped (a) and (b),  and  aligned (c) and (d). The most stable configuration, for this position,  is an antiferromagnetic state,
(a) and (b),  with  Mn-Mn distance equal 2.00~{\AA}. This small distance between Mn atoms 
lead to a direct overlap and consequently to an increase of the energy of the FM state when compared
to AFM configuration due to the Pauli exclusion principle.

\section{Conclusion}

In summary, we have performed a systematic study, using total
energy {\it ab initio} calculations, of the electronic and magnetic properties of Mn doped
Ge nanowire. Our results show that Mn has lower  formation energy at the center position of the GeNWs when compared to regions close to the surface. This indicates that once Mn atoms are placed inside the 
NWs, there will be an energetic barrier opposing its migration towards the surface.
However, if the NWs are saturated with H atoms, we expect a large concentration of Mn atoms at the surface, where the formation energy is lower than at the center of the wire by 0.6~eV.
The Mn impurity introduces a local magnetic moment due to $d$-levels resonant  with the VB. 
The Mn-Mn coupling is always ferromagnetic  except for first neighbor configurations.

We acknowledge support from the Brazilian agencies FAPESP and CNPq and
the computational facilities of the Centro Nacional de Processamento de Alto Desempenho
(CENAPAD-Campinas).
We thank  Dr. G. M. Dalpian for useful discussions.

%Email address: $^{*}$ajrsilva@if.usp.br; $^{**}$fazzio@if.usp.br


\begin{references}
\bibitem{extra01}
 For this Mn-Mn distance we use a doubled cell with diameter $d~\simeq~27.0$~{\AA} described in the
 calculation procedure with a Brillouin zone zampling  at the $\Gamma$-point.
\end{references}
\end{document}